# An Overview of Fingerprint-Based Authentication: Liveness Detection and Beyond

Filipp Demenschonok
University of Washington

Jason Harrigan
University of Washington

Tamara Bonaci[*]
Northeastern University
University of Washington

In this paper, we provide an overview of fingerprint sensing methods used for authentication. We analyze the current fingerprint sensing technologies, from algorithmic, as well as from hardware perspectives. We then focus on methods to detect **physical liveness**, defined as techniques that can be used to ensure that a *living* human user is attempting to authenticate on a system. We analyze how effective these methods are at preventing attacks where a malicious entity tries to trick a fingerprint-based authentication system to accept a fake finger as a real one (spoofing attacks). We then identify broader attack points against biometric data, such as fingerprints. Finally, we propose novel measures to protect fingerprint data.

## Introduction

Fingerprint sensing in consumer products is a ubiquitous technology. While the first major consumer product to include a finger-print sensor appeared on the market less than eight years ago (iPhone5S, launched in October 2013), it is projected that in 2020. there will be over 1 billion smart devices with fingerprint sensors.[1]

Today, fingerprint sensors are mainly used for authentication, as an alternative, or in addition to alpha-numerical passwords on a variety of mobile and smart devices. The drawback to the current finger-print-based authentication, however, is the fact that many existing systems are vulnerable to spoofing attacks, where an attacker tricks an authentication system to accept a fake fingerprint as a real one.[2]

Many successful spoofing attacks require only a minimal effort, for example an image, or a 3D model of a finger.[2] To make matters worse, it is relatively easy for a determined attacker to obtain victims' fingerprints, for example from a high resolution photograph of a finger, or even from fingerprints left on a device itself.[3]

Preventing spoofing attacks on fingerprint-based authentication systems has been an active research topic, and different approaches have been proposed, e.g., [4]. One of the proposed approaches is known as **the liveness detection**, and it involves detecting whether a live finger is being scanned on the device, as opposed to an image or artificial model of a finger.

In this paper, we present the current state of the art of the fingerprint-based authentication. We then analyze various existing and upcoming liveness detection methods, and investigate their potential for effective prevention of spoofing attacks. We further broaden our security analysis to a fingerprint-based authentication system as a whole, and identify several new possible vulnerabilities, and attack vectors against fingerprint data. In doing so, we focus on the flow of biometric data within the system, and analyze an impact that each point of compromise would have on a system as a whole, as well as on the biometric data, as a mean of authentication. In doing so, we analyze three new families of attacks against fingerprint-based authentication systems. Lastly, we propose a set of recommendation to prevent and mitigate the identified attacks. We note that, while we focus on mobile devices, and consumer platforms as a use case, this work is applicable to all platforms, using finger-print-sensing technology for authentication.

---

[*] **Corresponding author**, email: t.bonaci@northeastern.edu

To the best of our knowledge, we are the first to present the state of the art of fingerprint sensing methods, as well as anti-spoofing and liveness detection methods, which can be used concurrently with these methods. This paper aims to provide a novel, secure and holistic approach to the design of a fingerprint-based biometric system; including sensor selection, anti-spoofing and liveness detection defense strategies, and comprehensive information flow considerations through the system.

## System Security Context

### System

In this paper, our systems of interest are mobile and smart devices. Such systems usually have a camera, a microphone, wireless connectivity (usually both Wi-Fi and cellular), and they are typically connected to various cloud resources, such as social networks and financial services. In order to assure that only a device owner (and/or an authorized user) has access to the device's resources, the device is typically locked, and can be unlocked only by an authorized user. On a device equipped with a fingerprint sensor, such a sensor allows an authorized user to gain access to the system without having to type in a password.

### Assets

Providing access to a large number of resources and services, mobile devices are increasingly being considered personal and sensitive. As such, they consist of large amounts of information that a user is likely to want to protect. In many instances, such sensitive data may include: address book of contacts, emails, instant messaging history, pictures, location in-formation, users' activity and health information, users' financial information, as well as their purchasing history.

In addition, many users of mobile devices have mobile phone payment systems setup on their devices. These services provide an easy and convenient way to directly extract money from a compromised phone.

Lastly, under specific circumstances, a user might also care about protecting the bandwidth available to the smart device, its IP and MAC addresses, information about private network access, as well as computing resources of the device itself, as those could all be leveraged for a variety of nefarious purposes.

### Attackers

With such a wide range of assets to protect, there are a number of curious, or malicious entities that may want to attack the system. Based on the potential harm that they are likely to cause, those attackers can be grouped into **personal acquaintances**, **criminals**, and **potentially state actors**. While personal acquaintance may be interested in snooping through someone's private information (for example, social media accounts, messages and potentially, photographs), an access to the data stored on a device may allow criminals to blackmail a user, steal their financial assets, or even their identity.

A necessary requirement to prevent many of these attacks is user authentication, and one, frequently used, way to do so is fingerprint-based biometric authentication, which we discuss in the next Section.

## Fingerprint-based Authentication

All fingerprint sensor technologies rely, at least partially, on imaging the surface features of the human finger. It is known that the surface features of a human figure become permanent by the 16th week of pregnancy, and that those features are unique to an individual.[5] The unique features of the fingerprint minutia can be divided into three main categories:[6]

- **Level 1** features focus on the global pattern of the minutiae. These features are broadly characterized as arch, loop and whorl.
- **Level 2** features contain the ridges and values that make up the minutiae. The minutiae have unique features such as bifurcations or abrupt endings.
- **Level 3** features are all features that appear intra ridge. These include small details, such as sweat pores or warts/scars that deform the ridge shapes.

Most fingerprint-based algorithms authenticate an individual by identifying and matching their unique features. In doing so, the algorithms rely on two facts: **the persistence of fingerprints** – the fingerprints do not change over a person's lifetime (other than through external damage), and **the uniqueness of fingerprints**.[6]

### Fingerprint Matching Algorithms

Research and development of fingerprint matching algorithms have been a prolific research area, both in academia (see, e.g.,[7, 8]), and in consumer-based industries such as cellphones, computers and payment terminals.

Broadly speaking, a fingerprint matching algorithm operates by focusing on, and matching various properties of a fingerprint. Some example algorithms include: **minutia matching**, where the focus is on the level 2 features of a fingerprint,[7] **ridge flow maps**, where the focus is on extracting curves that correspond to the level 1 features of the fingerprint,[8] and **ridge wavelengths gradients**, where the focus in on assessing the ridge-to-ridge or ridge-to-valley spacing and orientation.[9] Each of these techniques can be used on their own, or in conjunction with each other. New algorithms continue to be developed to improve sensor performance, to overcome the limited minutia detail present on poor scans or small area sensors. The goal of all these techniques is to transform an arbitrary grey-scale image into some abstract form that allows for numerical comparison. The algorithm then sets a threshold for how stringent acceptance and rejection criteria should be, based on the features that it is monitoring.

The performance of a general authentication system is typically characterized by a **False Acceptance Rate (FAR) and** a **False Rejection Rate (FRR).** The FAR is a measure of how often an attacker is able to authenticate on the system because his/her fingerprint is similar enough to the users. The FRR represents a measure of how often a valid user must make multiple authentication attempts in order for the system to authenticate them. The FRR being high is a result of the system, and how the user interacts with it. If a user presses too hard, or lifts their finger up too fast, that could lead an otherwise good login attempt to be rejected, because the image captured was not good enough for the system to process. A low FRR ensures a good user experience.

The FAR and FRR present a natural tradeoff, as increasing security may make it more difficult for users to log in, while making the system too liberal with authentication attempts may reduce security. Sometimes both of these are reported as a single **Equal Error Rate (EER) statistic.** This number represents a point where the proportion of false matches is the same as the proportion of false non-matches and can be used as a single figure of merit to compare solutions.

Figure 1 (FAR vs. FRR) shows the distribution for the FAR and FRR cumulative distribution functions of a theoretical system. It can be seen from the image how selecting a certain threshold on the x-axis will result in different FAR and FRR performance of the system. Setting the threshold low will result in a low FRR, and a great user experience, but will compromise security. Setting a high threshold will result in a high security, but may become unusable in real world scenarios. A tradeoff between security and user experience will need to be set according to the intended application.

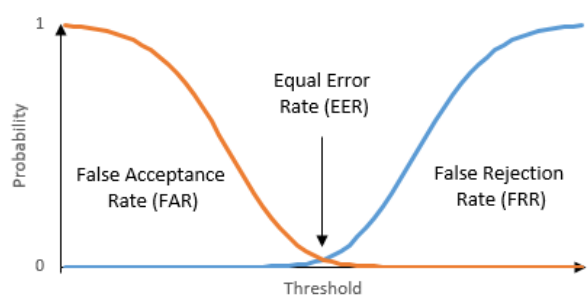

*Figure 1: FAR vs FRR - Selecting a certain threshold on the x-axis, will result in different FAR and FRR performance of the system*

## Physical Mechanisms of Fingerprint Detection

An important component of every fingerprint-based authentication system is a physical mechanism used to read a fingerprint. Based on the physical properties, we distinguish between four possible mechanisms:
- Optical mechanism,
- Capacitive mechanism
- Thermal mechanism, and
- Ultrasonic mechanism

### Optical Mechanisms

Optical fingerprint sensors are perhaps the oldest technology in this space. This category covers various Charge-Coupled Device (CCD) and CMOS implementations that capture the raw fingerprint image using a camera type sensor. These implementations typically rely on a prism that is flooded with diffused light. The light allows a sensor to capture an image of the fingerprint ridges.

The main advantage of these sensors comes from their ability to work through a thick mechanical stack-up above the sensing surface. This allows them to be implemented under cover glass on portable devices where other sensors cannot be used. Optical sensors, how-ever, suffer from multiple drawbacks. Older models struggle with detecting depth information and can be easy to spoof. The mechanical complexity of the optical assembly makes them space-prohibitive in many consumer electronics. These sensors also require significant height, as reducing the space between the prism and a sensor would introduce too much optical distortion.[10]

### Capacitive Mechanisms

Capacitive fingerprint sensors create an image of a fingerprint using the difference in capacitive coupling between the ridges and grooves of the skin. The benefit of this technology is that it does not require lenses, cameras, or an optical system, which allows it to be smaller than a traditional optical fingerprint scanner. Some drawbacks of capacitive technology are: the limits of dielectric constant and structure for materials that can be put over the sensor and still produce a good image, cost of large silicon area, and Electrostatic Discharge (ESD) vulnerability. This limits the thickness of protective material coatings that can be used and potential placements for these sensors.
There are two types of capacitive fingerprint sensing: passive and active.

#### *Passive capacitive fingerprint sensor*

The passive method (also called a direct method) of capacitive fingerprint sensing uses an open-ended circuit where a capacitive plate is exposed on the sensor surface, and the capacitance is continuously measured by a circuit. When the dielectric constant of skin is sensed to be present, due to the change in capacitance, the outline of the skin ridges can be detected.

#### *Active capacitive fingerprint sensor*

An active capacitive fingerprint sensor conducts low frequency RF energy into the skin and receives the signal in each of the capacitive elements, such that a closed loop circuit is formed. The sensors are composed of an array of sensing plates that can detect this signal when it is coupled through the skin into the sensor. Typically, this approach is more sensitive than the passive method but requires additional circuitry to work.

### Thermal Mechanisms

Thermal fingerprint sensors have been proposed as a lower cost, and more robust alternative to capacitive sensors. The operating principle is using materials that convert changes in temperature to a voltage.
One issue in general with thermal measurement methods is that the temperature quickly equalizes between ridges and grooves of the finger once applied to the sensor, which means that the thermal image is only available for a short period of time, and quickly disappears. This makes passive thermal imaging infeasible. However, a voltage signal can be applied to the thermo-junction, and a temperature signal can be generated which heats up and cools off dependent on the heat capacity of the material contacting it. This allows an image of a fingerprint to be sensed. Because it requires active heating, the temperature-based fingerprint sensor has higher power requirements than capacitive sensors.

### Ultrasonic Mechanisms

One of the newest and most promising technologies emerging in the fingerprint sensor space are ultrasonic micro-electromechanical systems (MEMS) type sensors.[11] These sensors rely on array of microsismic sensors similar to those employed in the ultrasound medical industry. These transducers work as both the receiver and transmitter of the signal. First a pulse is sent out, and then the sensors monitor their own mechanical deformation to determine the mechanical properties of the material placed on them. The resonant frequency of each "pixel" is affected by being highly damped when a ridge is pressed or being undamped when the air from a valley is present. With multiple images and various image processing techniques these sensors are able to build a 3-D image of the finger. In addition to relying on the surface minutia additional biometric markers from the dermis, such as vein structure can be added to the model. This makes this type of sensor particularly difficult to spoof as any artifact used would need to have detailed reproductions of the internals of a victim's finger.[11]

## Spoofing Attacks and Prevention Through Liveness Detection

While convenient to use, many of the existing fingerprint-based authentication system are vulnerable to spoofing attacks, where an attacker tricks an authentication system to accept a fake fingerprint as a real one.[2]

To prevent an onslaught of spoofing attacks, many fingerprint-based authentication methods additionally employ liveness detection. Liveness detection refers to any technique that can be used to ensure that a true, living human user is attempting to authenticate on the system.[12] Several such methods are currently being used in consumer devices, and many more are under development. In the rest of this section, we provide an overview of several such liveness detection methods.

### Skin Temperature Detection

One of the simplest (and minimal effort) ways to detect liveness is by sensing the temperature of the finger. While being low-cost and easy to implement, this detection method is also easy to bypass by simply warming up the fake finger to body temperature. Due to the possible variation in temperature between different people's fingertips, including in cold weather, the limits cannot be too stringent. However, temperature detection can be an additional easy check to help augment a more comprehensive liveness detection system.

### Heartbeat Detection

Another simple method to detect a spoofing attack against a fingerprint-based authentication system is to check for the presence of a heartbeat. If no heartbeat is present, or if an unrealistic value is obtained, the authentication attempt can be ruled out as a spoof. One way to detect a heartbeat can be achieved by measuring the amount of light that is transmit-ted through a finger. In doing so, we rely on the fact that hemoglobin has a differential absorbance, at particular wavelengths, as a function of its oxygen saturation.[12] By monitoring the change in the absorption, an estimate of the heartbeat can be obtained.

### Finger Translucence

Live tissue within the finger has the capability of passing light of certain wavelengths in a characteristic way, while partially filtering out other wavelengths. Relying on that fact, an-other method of liveness detection uses a white LED to shine light into the finger at one location, then uses one or more photodiodes to detect light spectrum passed through the finger to another location. A fake finger may be opaque and not pass any light, or it may pass all wavelengths of light equally or in a way that a live finger would not.

### Skin Impedance

Measuring the resistance of skin has also been investigated as a means of liveness detection. Various factors including humidity and moisture of the finger affect the surface resistance, leading to quite a wide range of potential resistances.[13] The impedance can also be measured at frequencies above DC. In general, this can be useful as a secondary means of liveness detection at best, due to the wide range of possible impedances between individuals and environmental conditions.

### Skin Perspiration Patterns

The way in which perspiration can indicate the presence of a live finger has also been studied as a way to detect liveness.[14] This method is based on the observed phenomena of perspiration traveling along finger ridges, starting from skin pores and extending outwards over time. This change over time (for example, over a period of 2 seconds) is only present in live humans, and not cadavers or spoofed fingerprints.

Furthermore, the perspiration pattern is unique to an individual's finger, and is determined by the location of sweat pores and surface texture of the finger. This is intriguing because not only can it function as a general detection of liveness, but it may also have the potential to be another signal to help identify individuals as well, with the added benefit that it may not be easy to forensically get this marker from fingerprints left on surfaces.

### Skin Deformation on Touch

Human flesh has multiple characteristics that are intrinsic to it, such as tensile strength, tensile strain, compressive strength, shear strain and hardness.[15] By taking multiple images during an authentication attempt, these properties can be estimated from the images obtained. If graphical analysis is not in line with nominal human values, the sensor can reject the login attempt. Other approaches focus on obtaining even more proactive measurements of the mechanical properties of the finger. By forcing the user to rotate the finger or move it about on the surface of the sensor, a more precise estimate of the mechanical properties can be made.[15] This method is particularly interesting because it works with any of the aforementioned fingerprint sensing technologies, and it would only require that the manufacturer characterizes nominal mechanical properties for human fingers on their sensors. The drawback of this approach, however, is that it requires fast sensors and processors to take multiple good images during a short login press. Alternatively forcing the user to rotate their fin-ger during device usage could be seen as unacceptable from an experience standpoint.

### Internal Imaging of Dermis

One of the primary advantages of ultrasonic sensors is the ability to image beyond the top layer of the epidermis. The skin on a finger is a complicated organ that contains several layers. The use of ultrasonic transducers allows the sensor to image these internal layers as well as vein/capillary structures and bone structure. The combination of these internal fingerprint characteristics makes a spoof nearly impossible to reproduce, as it would require a nearly perfect model of the victim's finger.

## Summary of fingerprint sensing methods, and liveness detection techniques

With respect to their ability to detect a spoofed finger, we classify liveness detection methods into three categories:
- Low security methods,
- Medium security methods, and
- High security methods,

where a low security method refers to a method that can be easily tricked to accept a spoofed finger and a high security method refers to method that would require a substantial amount of skill and effort from an attacker

Our results are summarized in Table 1.

*Table 1: Relative security of various liveness detection methods*

| | |
|---|---|
| Low security | Skin temperature |
| | Skin impedance |
| Medium security | Heartbeat detection |
| | Finger translucence |
| | Skin deformation |
| High security | Skin perspiration detection |
| | Internal imaging of capillaries |

As a general recommendation, we see capacitive fingerprint sensing method, with perspiration-based liveness detection, and ultrasonic sensing with capillary imaging-based liveness detection as two attractive combinations of fingerprint sensing and liveness detection technologies. We believe that these methods could improve the security by preventing finger-print-spoofing attacks in consumer devices without a negative impact on the devices' price point.

## Broader Analysis of Attacks Against Biometric-Based Authentication

One of the features of many biometric-based authentication methods, such as fingerprint sensors, is the fact that every person's fingerprint is unique and permanent, and as such cannot be reissued or updated, if they ever get compromised. This feature effectively means that protecting biometric data, during the collection phase, as well as during the authentication phase is a critical task. Compromised fingerprint data would lead to a compromise of that specific authentication system, but also to a compromise of all existing and future systems that do (would) rely on the fingerprint data for authentication.

These unique challenges of fingerprint data; the critical need to protect fingerprint data, and the difficulty of protecting the data using traditional cryptographic methods, require careful consideration. We start this process by observing that every fingerprint-based authentication system has a specific information flow, consisting of three phases:

- **Enrollment phase** – a process during which a user of the system allows their fingerprint data, referred to as a fingerprint template, to be collected, and stored by the system. This step typically occurs only once.
- **Storage phase** – all times during which the collected data is not being used.
- **Authentication phase** – a process during which the stored fingerprint template is compared with the fingerprint data collected at that moment, in order to identify, and authenticate the user.

We focus on each of the defined steps of the information flow (enrollment, storage and authentication), and recognize that those phases can be targeted by specific attacks, as depicted in Figure 2 (Attacker Model):

1. Physical attacks,
2. Sensor-to-host Man-in-the-Middle (MitM) attacks,
3. Host-to-Server MitM attacks, and
4. Local and remote attacks.

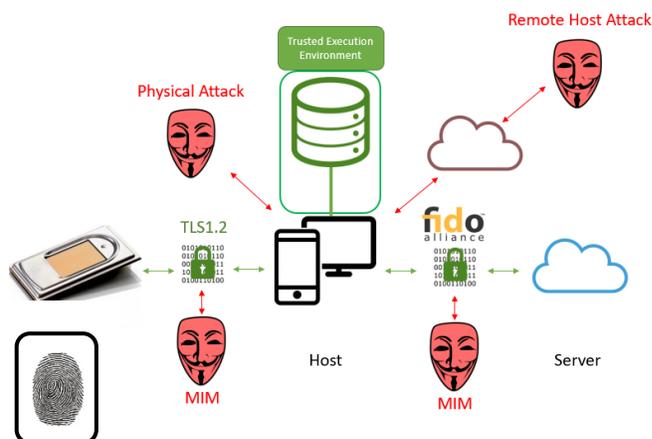

*Figure 2 Attacker model*

**Note:** in this paper, we focus on data obtained using fingerprint sensors, but the challenges discussed here apply equally to other forms of biometric authentication that rely on data recorded during the enrollment phase, and later used for authentication.

Starting from the assumption that the best security practices have been followed in ensuring physical security of a fingerprint sensor, as well as of a smart (mobile) device, and the potential cloud storage, in the rest of the section we focus on attack classes (2)-(4).

## Sensor-to-Host Man in the Middle

The first attack vector we consider is a Man-in-the-Middle (MitM) attack, mounted against the data *en route* between the sensor and the host processor. These attacks become feasible by physically compromising the data bus, or by picking up (eavesdropping on) the wireless communication on the bus. The first step to avoiding this class of attacks is encrypting the communication between the host and the sensor, in order to ensure that the fingerprint data is never passed in plaintext, and not compromised (neither during enrollment, nor during authentication phases). If the device is monolithic, this can be achieved in the factory by providing a secret session key to both endpoints during manufacture. If this is not possible, or the system is modular, a TLS or similar type of scheme can be used. An asymmetric key encryption algorithm can be used to establish a session key, and then a symmetric key algorithm can be used to encrypt the data.

### Host-to-Server Man in the Middle

The second opportunity for a MitM attack is between the host and any external resources, such as a cloud server, where the fingerprint data may be stored. These types of attack can be carried out by intercepting traffic between two endpoints. A form of encryption and authentication is required to ensure that communication is not compromised. The Fast Identity Online Alliance (FIDO) is a non-profit organization that has codified a consistent way that biometric credentials can be shared across the internet. FIDO provides a password-less, and second factor authentication user experience. On the user end of the system, users are able to use their fingerprint sensor without any additional steps. In the background, however, FIDO uses an asymmetric encryption algorithm, and locally on the host, associates bio-metric authentication with the release of a public key to the remote resource. Challenges are passed using the private key to establish a secure session.

### Local and Remote Host Attacks

In both local and remote attacks, an attacker's goal is to load malware onto the system, in order to gain access to privileged information. In doing so, an attacker can gain a privileged access to system memory, or they can again intercept outgoing/incoming data transmissions.

Local host attacks would require the attacker to somehow manipulate the host, so as to load malware on to it. One of the easiest, and most common methods would be through an USB port. Remote host attacks, on the other hand, would use a different vector to load malware on to the system. This would be achieved with phishing, infected peer-to-peer clients, infected emails, or even through social engineering.

There exist many widely known techniques to achieve the malicious goals. An attack, such as PoisonTap through a fake USB device, could forward outgoing Ethernet data to the at-tacker. A remote attack, such as row hammering, could potentially gain access to system memory.

The best way to prevent and mitigate these attacks is to implement a solution that stores the fingerprint templates, as well as incoming sensor data in a Trusted Execution Environment (TEE). One such approach is the Intel SXG enclave environment, which ensures that the system remains protected even when the BIOS, VMM, OS, and drivers are compromised. In mobile environments that use ARM based processors, the ARM® TrustZone® is able to achieve a similar goal.

## Discussion: Template Storage Design Choices

The applications for fingerprint sensors, and biometric data as a whole are increasingly being used in a variety of applications. This is evident in the implementation of biometric passports, biometric driver's licenses, as well as in a variety of commercial applications, all of which rely on fingerprint data for authentication. This shift in a paradigm will leave designers and system architects with a variety of options, that may have non-trivial, even serious consequences. In this paper, we assumed that the template information is stored in secure memory on the host device, **but such an approach may not be preferable, or even feasible in certain applications.**

On one end of the spectrum, the template database could be implemented on the sensor silicon itself. In this scenario, only a pass or fail bit would need to be sent to the host, ensuring that even a compromised host would never leak private biometric data. The downside of this approach is that such a system would require an additional embedded storage, and computing resources on the sensor, which would come at additional cost, and likely per-form worse than the comprehensive resources available on a typical host system. An alternative approach altogether could see the template data stored in the cloud, or on remote servers. Such an approach could allow for a user to access multiple devices, or terminals on the same domain without having to explicitly enroll on each device. This can have usability advantages in settings where many users access and share public resources, for example, meeting rooms with smart boards or computer-controlled projectors. This approach, however, adds important security questions to the system. For example:

- Would users trust third parties, e.g., employers, to store and protect their biometric data?
- Will sensor manufacturing variation allow for an enroll template made on one sensor to authenticate on another copy of the same sensor?
- Would such templates be comprehensive enough to be used across different sensors from different manufacturers?

The reality is that the implementation of the template data storage is left to the creator of the biometric algorithms. The end user does not know where and how is their data being stored.

In the best-case scenario, users' data would be stored in a format abstracted and obfuscated enough to make it unusable for an attacker able to compromise template database. In the worst-case scenario, the data may end up being stored with enough detailed information to create a spoof if the database is com-promised. As discussed earlier, that represents a critical problem for fingerprint-based authentication system, because once the fingerprint template is compromised, it may compromise all other systems as well.

Aside from that security concern, biometric data can provide sensitive information about a user, such as information about their gender or race. These concerns raise additional questions: do possible privacy challenges imply that biometric data requires an even more stringent protection? Moreover, should developers be under a legal obligation to disclose to the end user how the biometric data is stored, and how it is processed in their algorithms?

## Conclusion

Fingerprint sensing is rather ubiquitous in consumer and mobile device space as a way to authenticate users (in lieu, or in addition to alphanumerical passwords). Fingerprint sensors of today, however, are susceptible to a variety of possible attack vectors. One of the easiest, and most prevalent attacks against fingerprint sensors is a fingerprint spoofing attack. In this attack, fake fingers, 3D models of fingerprints, or potentially even a severed finger are used to trick an authentication system into accepting a malicious entity as a valid user. The next generation of fingerprint-enabled devices will likely include robust liveness detection methods as a way to prevent such spoofing attacks.

In this paper, we present a state-of-the-are review of the existing fingerprint sensing method, used in the modern-day consumer devices. We then provide a detailed analysis of existing, and upcoming liveness detection methods. In doing so, we observe that most liveness detection methods will deter, and stop opportunistic criminals, but may fail at stopping a determined attacker. There do, however, exist several liveness methods that would be hard to bypass or compromise, regardless of an attacker's effort. The two most promising methods are internal capillary imaging and ultrasonic based sensors.

Widening our analysis of possible attacks against fingerprint-based authentication methods, we next make two important observations; that fingerprint data are unique and permanently tied to a user, but their samples (slightly) vary from one measurement to another. These two observations make the needed to protect the biometric data critical for the security of all fingerprint-based authentication systems. At the same time, these observations render many of the traditional security (cryptographic) methods of protecting data infeasible.

Focusing on the flow of information in a typical biometric-based authentication system, we next identify several new families of attack vectors. We analyze each of these families, and make recommendation as to how to protect the biometric data during each of the phases of information flow.

A broader goal of our paper is to emphasize the importance of protecting biometric data, especially in the consumer space. While in this paper, we focus on the fingerprint data, many of the issues raised translate to other biometric data as well. Their unique, and permanent tie to a user makes the need to protect such data critical, and not just from the security standpoint, but also from privacy.